\documentclass[fleqn,twoside]{article}
\usepackage[headings]{espcrc2}

\readRCS
$Id: espcrc2.tex,v 1.2 2004/02/24 11:22:11 spepping Exp $
\usepackage{graphicx}
\usepackage[figuresright]{rotating}

\usepackage{epsfig}
\usepackage{dcolumn} 
\usepackage{bm} 
\usepackage{latexsym}
\input {colordvi}
\def\lsim{\raise0.3ex\hbox{$<$\kern-0.75em\raise-1.1ex\hbox{$\sim$}}}
\def\gsim{\raise0.3ex\hbox{$>$\kern-0.75em\raise-1.1ex\hbox{$\sim$}}}

\def\mean#1{\left<#1\right>}
\def\Journal#1#2#3#4{{#1}{\ #2} (#4) {#3} }

\def\JPG{{J. Phys}~{G}}

\def\NPA{{Nucl. Phys. A}}
\def\NPB{{Nucl. Phys. B}}
\def\PLB{{Phys. Lett. B}}

\def\PRL{Phys. Rev. Lett.\ }
\def\PR{{Phys. Rev.}}
\def\PRD{{Phys. Rev. D}}
\def\PRC{{Phys. Rev. C}}

\def\ZPC{{Z. Phys. C}}
\def\ARNPS{{Ann. Rev. Nucl. Part. Sci.\ }} 

\newcommand{\AmS}{{\protect\the\textfont2
  A\kern-.1667em\lower.5ex\hbox{M}\kern-.125emS}}

\title{Measurements of Hard-Scattering by PHENIX at RHIC}

\author{M.~J.~Tannenbaum\address[MJT]{Brookhaven National Laboratory,\\ 
Upton, NY 11973-5000 USA}%
        \thanks{Research supported by U.S. Department of Energy, DE-AC02-98CH10886.},
        {for the PHENIX Collaboration.}}
       
\runtitle{Measurements of Hard-Scattering by PHENIX at RHIC}
\runauthor{M.~J.~Tannenbaum}

\begin{document}

\begin{abstract}
Hard-scattering in p-p collisions was discovered in 1972 at the CERN-ISR, the first hadron collider. Techniques were developed and several hard-processes were discovered which form the basis for many of the measurements made in p-p and Au+Au collisions at RHIC. Recent measurements of hard-scattering and related reactions by the PHENIX experiment at RHIC are presented in this context. \vspace{1pc}
\end{abstract}

\maketitle

\section{INTRODUCTION---METHODS FROM THE CERN-ISR}
\label{sec:intro}
 The CERN-ISR was the first Hadron Collider and made many discoveries~\cite{JK1,GJ1} and developed many techniques which are in use in Au+Au and p-p collisions at RHIC today:
 \begin{itemize}
 \item The rapidity plateau;
 \item Hard-scattering in p-p collisions via particle production at large $p_T$ (see Fig.~\ref{fig:I1}) which proved that the partons of Deeply Inelastic Scattering interacted strongly with each other; 
 \item $x_T$ scaling measurements to find the underlying physics.
 \item Proof using same-side and away-side two particle correlations that high $p_T$ particles in p-p collisions are produced from states with two roughly back-to-back jets which are the result of scattering of the constituents of the nucleons as described by QCD which was developed during the course of these measurements.
 \item Direct lepton ($e^{\pm}$) production from the decay of (unknown at that time--1974) particles composed of $c$ and $b$ quarks (see Fig.~\ref{fig:I2}).
 \item First and only $J/\Psi$ cross section measurement integrated over all $p_T\geq 0$ at a hadron collider until PHENIX at RHIC~\cite{PXJPsi1} and CDF~\cite{CDFJPsi} (15 years after their first publication).
 \item Direct Photon Production.
 \end{itemize}
 Some of these results are reviewed in Figs.~\ref{fig:I1} through \ref{fig:I4}.
\subsection{Discovery of High $p_T$ $\pi^0$ production}
\begin{figure}[!h]
\vspace*{-1.5pc}
\begin{center}
\includegraphics[width=0.75\linewidth]{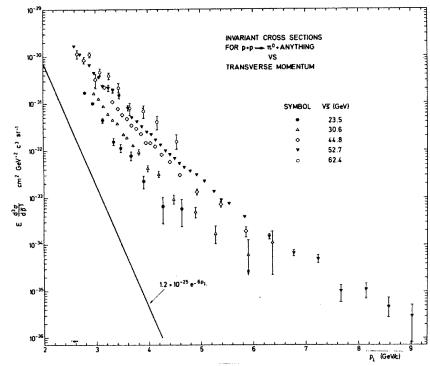}
\end{center}\vspace*{-3pc}
\caption[]{Invariant cross section of $\pi^0$ vs. $p_T$ at mid-rapidity in p-p collisons for 5 values of $\sqrt{s}$~\cite{CCR}. }
\label{fig:I1}\vspace*{-1.2pc}
\end{figure} 

Fig.~\ref{fig:I1} shows the first measurement of $\pi^0$ production at large $p_T$. The $e^{-6 p_T}$ dependence at low $p_T$ breaks to a power law with characteristic $\sqrt{s}$ dependence for $p_T\geq 2$ GeV/c.
\subsection{Discovery of direct $e^{\pm}$}

\begin{figure}[!h]
\vspace*{-2.5pc}
\begin{center}
\includegraphics[width=1.13\linewidth]{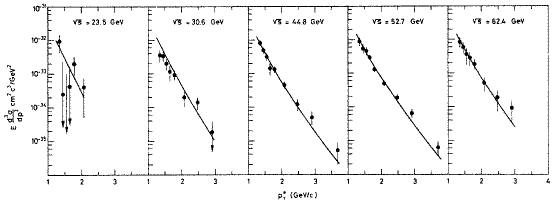}
\end{center}
\vspace*{-3pc}
\caption[]{Invariant cross sections at mid-rapidity: $(e^+ + e^-)/2$ (points); $10^{-4}\times (\pi^+ +\pi^-)/2$ (lines)~\cite{CCRS}.}
\label{fig:I2}
\end{figure}

Fig.~\ref{fig:I2} shows the $p_T$ spectra of direct $e^{\pm}$ for all 5 values of $\sqrt{s}$ at the ISR~\cite{CCRS}, reasonably described by the relationship $(e^+ + e^-)/(\pi^+ + \pi^-)\approx 10^{-4}$. The internal conversion of direct photons~\cite{FarrarFrautschi} at a level $\gamma/\pi^0\sim 10-20\%$ for $p_T\geq 1.3$ GeV/c was originally proposed as the source of the direct leptons, but CCRS~\cite{CCRS} looked and set limits excluding this. After the discovery of charm particles two years later in 1976, it was shown that the direct $e^{\pm}$ were due to the semi-leptonic decay of charm mesons~\cite{HLLS}. 
\vspace*{-2pc} 
\subsection{$J/\Psi$ Measurements} 
\begin{figure}[!h]
\vspace*{-0.5pc}
\begin{center}
\begin{tabular}{cc}
\includegraphics[width=0.48\linewidth]{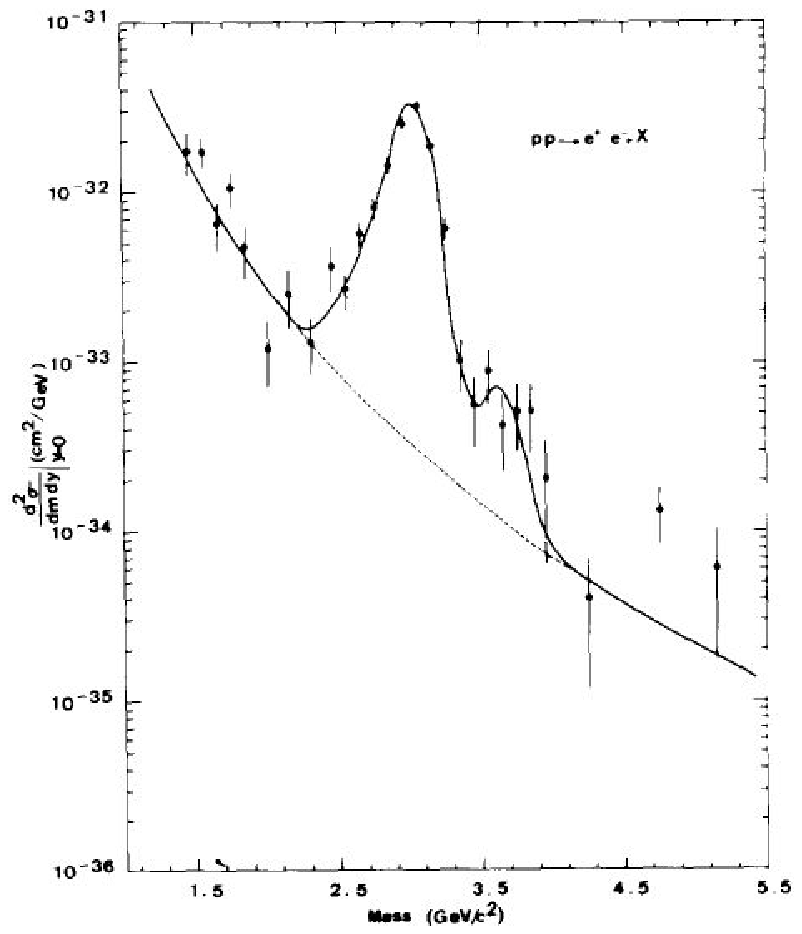}\hspace*{-0.04\linewidth}&
\includegraphics[width=0.48\linewidth]{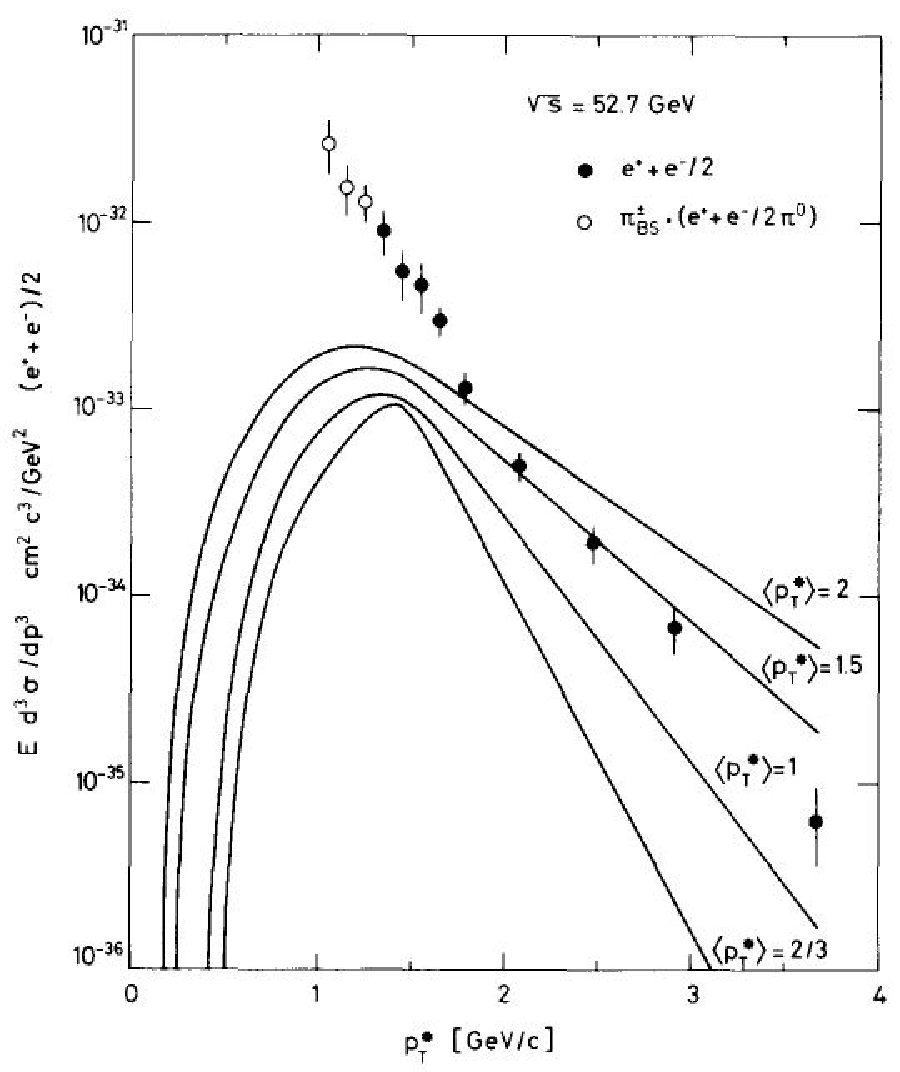} 
\end{tabular}
\end{center}
\vspace*{-3pc}
\caption[]{(left) $d\sigma_{ee}/dm_{ee} dy|_{y=0}$~\cite{Clark}; (right) $e^{\pm}$ data at $\sqrt{s}=52.7$ GeV (Fig.~\ref{fig:I2}) with calculated $e^{\pm}$ spectrum for $J/\Psi$ for several values of $\mean{p_T}$~\cite{CCRS}. }
\label{fig:I3}
\end{figure}

Fig.~\ref{fig:I3}-(left) shows the best $J/\Psi$ measurement at the ISR~\cite{Clark} while Fig.~\ref{fig:I3}-(right)~\cite{CCRS} shows that the direct electrons (Fig.~\ref{fig:I2}) are not the result of $J/\Psi$ decay since $\mean{ p_T}=1.1\pm 0.05$ GeV/c~\cite{Clark}.  
\subsection{Discovery of Direct photons and measurements of correlations}
\begin{figure}[!h]
\begin{center}\vspace*{-1.5pc}
\begin{tabular}{cc}
\includegraphics[width=0.48\linewidth,height=0.55\linewidth]{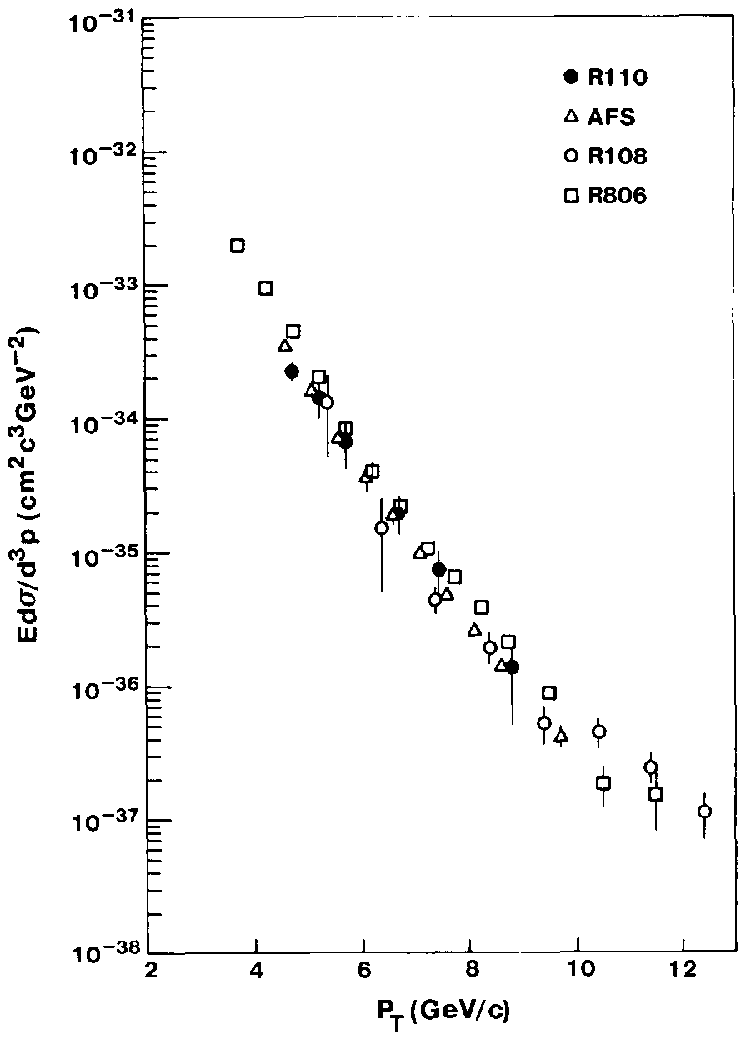}\hspace*{-0.04\linewidth}&  
\includegraphics[width=0.48\linewidth]{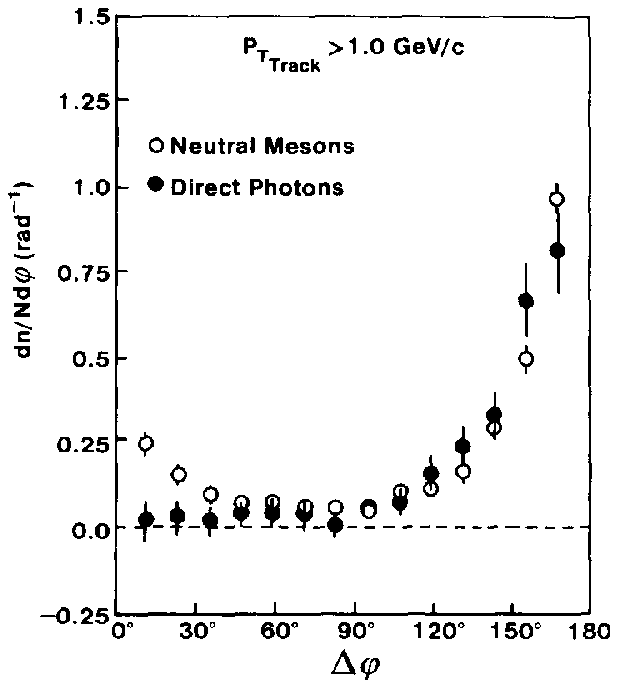} 
\end{tabular}
\end{center}\vspace*{-2pc}
\caption[]{(left) Compilation of invariant cross sections of direct-$\gamma$ production at ISR~\cite{CMOR}; (right) azimuthal correlations of neutral mesons and direct-$\gamma$ with $h^{\pm}$~\cite{CMOR}. }
\label{fig:I4}\vspace*{-2pc}
\end{figure}

Fig.~\ref{fig:I4}-(left) shows all direct photon measurements from the ISR~\cite{CMOR} while Fig.~\ref{fig:I4}-(right) shows the azimuthal correlation of charged tracks with $p_T \geq 1$ GeV/c to either a neutral meson or direct photon trigger with $p_{T_t}\geq 6$ GeV/c. The absence of same-side correlation with direct photons confirms that photons from fragmentation are not a significant source of direct photons in this measurement~\cite{CMOR}.
\section{MEASUREMENTS BY PHENIX AT RHIC IN P-P AND AU+AU COLLISIONS}
\begin{figure}[!ht]
\begin{center}
\begin{tabular}{cc}
\includegraphics[width=0.48\linewidth]{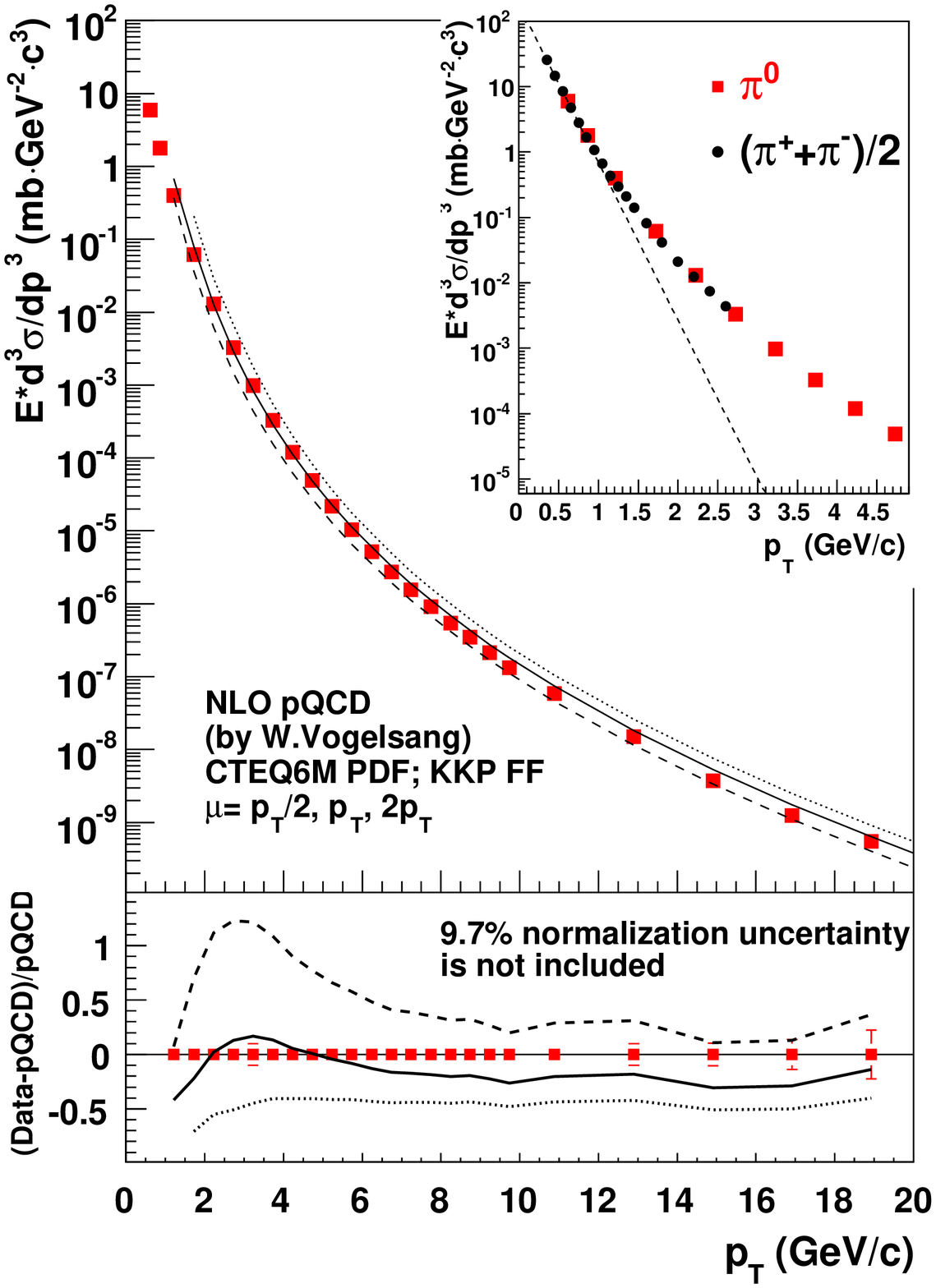} &
\hspace*{-0.04\linewidth}\includegraphics[width=0.48\linewidth]{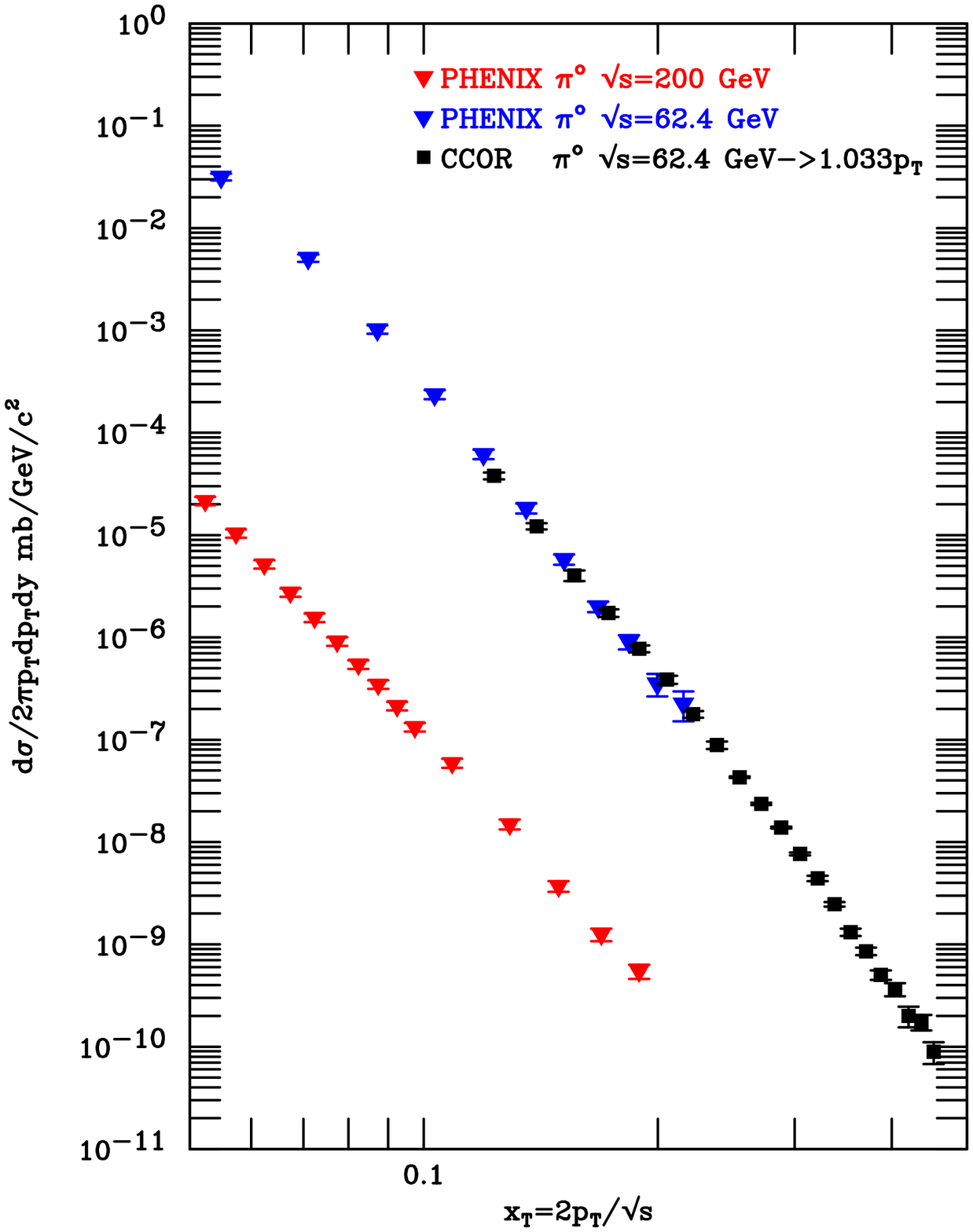}
\end{tabular}
\end{center}\vspace*{-2.5pc}
\caption[]{(left) $Ed^3\sigma/dp^3$ vs. $p_T$ for mid-rapidity $\pi^0$ at $\sqrt{s}=200$ GeV in p-p collisions~\cite{PXpi1}; (right) same data plus PHENIX and CCOR-ISR~\cite{CCOR} measurements at $\sqrt{s}=62.4$ GeV, where the absolute $p_T$ scale of the ISR measurement has been corrected upwards by 3\% to agree with the PHENIX data.} 
\label{fig:f1}\vspace*{-1.5pc}
\end{figure}

The high $p_T$ $\pi^0$ cross section in p-p collisions at RHIC provides a baseline for measurements in A+A collisions. Fig.~\ref{fig:f1}-(left) shows the the invariant cross section measured from the exponential range at low $p_T$ to the region of hard-scattering, the power law for $p_T\geq 2$ GeV/c, where the data agree with QCD~\cite{PXpi1}. The power law is emphasized by the log-log plot in Fig.~\ref{fig:f1}-(right) of PHENIX measurements of $\pi^0$ at $\sqrt{s}=200$ and 62.4 GeV and CCOR measurements~\cite{CCOR} at 62.4 GeV as a function of $x_T=2p_T/\sqrt{s}$. This plot exhibits that the cross section for hard-processes obeys the scaling law:
\begin{equation}
E{{ d^3\sigma} \over {d^3p}}={1 \over p_T^{\,n_{\rm eff}} } F ({p_T \over \sqrt{s} })={1 \over \sqrt{s}^{\,n_{\rm eff}} } G({x_T} ) 
\end{equation}
where $n_{\rm eff}(x_T,\sqrt{s})\sim 4-6$ gives the form of the force-law between constituents as predicted by QCD with non-scaling structure and fragmentation functions and running coupling constant~\cite{BBKBBGCGKS}. 
\subsection{Suppression of $\pi^0$ in Au+Au collisions} 
   Since hard-scattering at high $p_T >2$ GeV/c is point-like, with distance scale $1/p_T < 0.1$ fm, the cross section in p+A (A+A) collisions, compared to p-p, should be larger by the relative number of possible point-like encounters, a factor of $A$ ($A^2$) for p+A (A+A) minimum bias collisions. When the impact parameter or centrality of the collision is defined (typically by the measured multiplicity which increases roughly proportional to the number of nucleons which are struck in the reaction---called the number of participating nucleons,  or participants, $N_{\rm part}$) the proportionality factor is $\mean{T_{AA}}$, the average overlap integral of the nuclear thickness functions.

\begin{figure}[!ht]
\begin{center}
\begin{tabular}{cc}
\hspace*{-0.02\linewidth}\includegraphics[width=0.53\linewidth,height=0.4\linewidth]{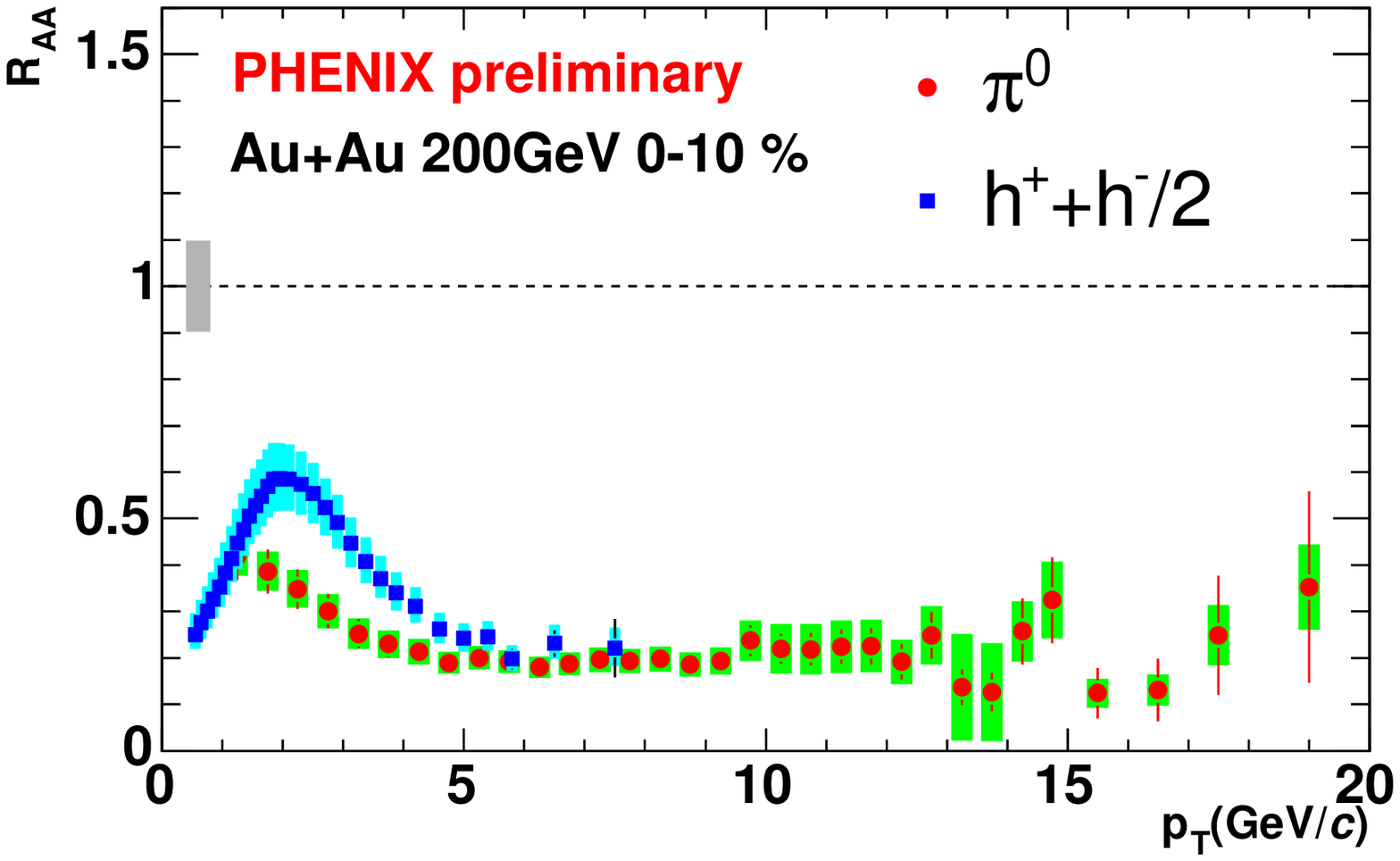} &\hspace*{-0.08\linewidth} 
\includegraphics[width=0.47\linewidth]{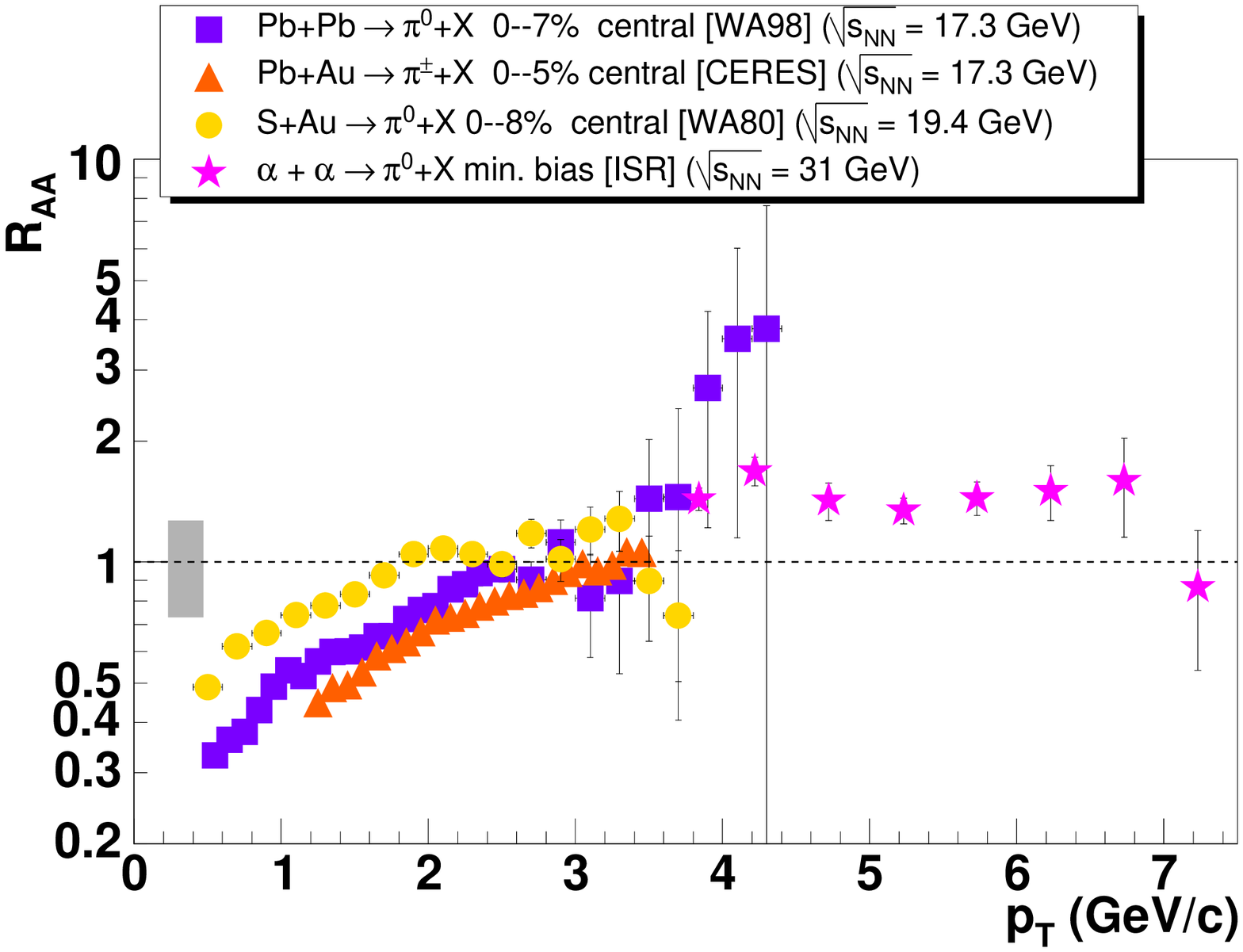} 
\end{tabular}
\end{center}\vspace*{-2.5pc}
\caption[]{(left) $R_{AA}(p_T)$ for $\pi^0$ and $h^{\pm}$ in central Au+Au collisions at $\sqrt{s_{NN}}=200$ GeV~\cite{pi0-QM05}; (right) compilation of $R^{\pi^0}_{AA}(p_T)$ at lower $\sqrt{s_{NN}}$~\cite{DdE}. }
\label{fig:f2}\vspace*{-1.5pc}
\end{figure}

   The discovery, at RHIC, that $\pi^0$ are suppressed by roughly a factor of 5 compared to point-like scaling of hard-scattering in central Au+Au collisions is arguably {\em the}  major discovery in Relativistic Heavy Ion Physics. In Fig.~\ref{fig:f2}-(left), the data for $\pi^0$ and non-identified charged particles ($h^{\pm}$) are presented as the ratio of the yield of $\pi^0$ per central Au+Au collision  (upper 10\%-ile of observed multiplicity) to the point-like-scaled p-p cross section:
   \begin{equation}
  R_{AA}(p_T)={{d^2N^{\pi}_{AA}/dp_T dy N_{AA}}\over {\langle T_{AA}\rangle d^2\sigma^{\pi}_{pp}/dp_T dy}} \quad . 
  \label{eq:RAA}
  \end{equation}
Since there is no suppression of $\pi^0$ in any measurement in A+A collisions at lower c.m. energies, $\sqrt{s_{NN}}\leq 31$ GeV/c (Fig.~\ref{fig:f2}-(right)), the suppression is unique at RHIC energies and occurs at both $\sqrt{s_{NN}}=200$ and 62.4 GeV. The suppression is attributed to energy-loss of the outgoing hard-scattered color-charged partons due to interactions in the presumably deconfined and thus color-charged medium produced in Au+Au (and Cu+Cu) collisions at RHIC~\cite{BSZARNPS}. Fig.~\ref{fig:f2}-(left) also shows that the suppression of non-identified charged hadrons and $\pi^0$ are different for $2\leq p_T \leq 6$ GeV/c. An $x_T$ scaling analysis showed that the $\pi^0$ in Au+Au central and peripheral collisions at $\sqrt{s_{NN}}=200$ and 130 GeV followed $x_T$ scaling with the same value of $n_{\rm eff}$ as in p-p colllisions suggesting that the energy loss scaled, i.e. was a constant fraction of the $p_T$ of the outgoing parton~\cite{PXxTAuAu}. The $h^{\pm}$ do not $x_T$ scale because baryons are not suppressed for $2\leq p_T\leq 6$ GeV/c~\cite{ppg015}. This is called the baryon-anomaly, which is still not understood. 
\subsection{$J/\Psi$ suppression}
\begin{figure}[!ht]
\begin{center}\vspace*{-1.5pc}
\begin{tabular}{cc}
\hspace*{-0.015\linewidth}\includegraphics[width=0.48\linewidth,height=0.6\linewidth]{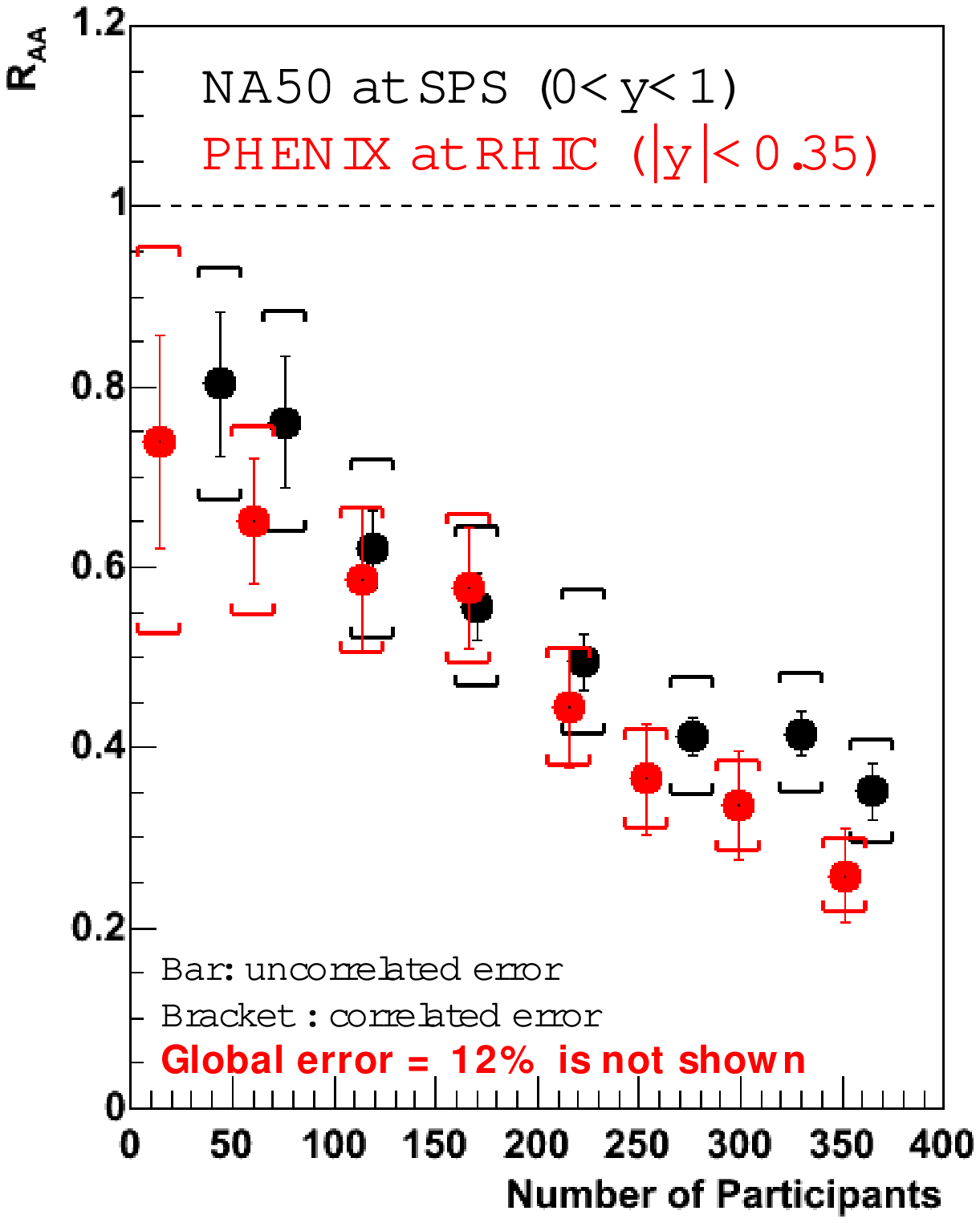} & \hspace*{-0.05\linewidth} 
\includegraphics[width=0.48\linewidth,height=0.6\linewidth]{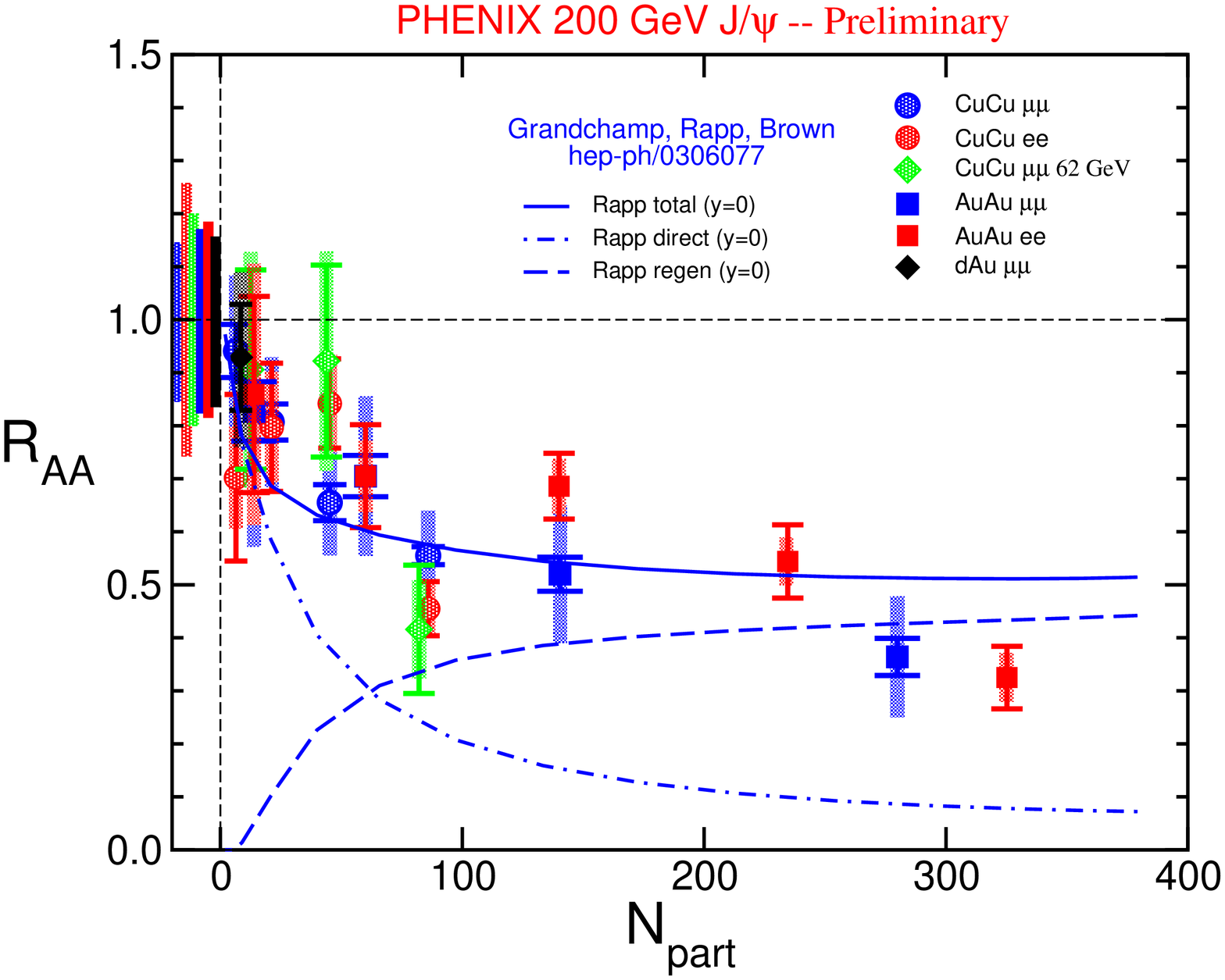}  
\end{tabular}
\end{center}\vspace*{-2.5pc}
\caption[]{(left) $R_{AA}^{J/\Psi}$ vs centrality ($N_{\rm part}$) at RHIC and SpS energies~\cite{PRL98-1}. (right) Predictions for $R_{AA}^{J/\Psi}$ in a model with regeneration~\cite{GRB}.}
\label{fig:f3}\vspace*{-2.5pc}
\end{figure}

The dramatic difference in suppression of hard-scattering at RHIC compared to  SPS fixed target c.m. energy ($\sqrt{s_{NN}}=17$ GeV) stands in stark contrast to $J/\Psi$ suppression, originally thought to be the gold-plated signature for deconfinement and the Quark Gluon Plasma (QGP)~\cite{MatsuiSatz}. $R_{AA}$ for $J/\Psi$ suppression is the same, if not identical, at SPS and RHIC (see Fig.~\ref{fig:f3}-(left)), thus casting a serious doubt on the value of $J/\Psi$ suppression as a probe of deconfinement. The medium at RHIC makes $\pi^0$'s  nearly vanish but leaves the $J/\Psi$ unchanged compared to lower $\sqrt{s_{NN}}$. One possible explanation is that $c$ and $\bar{c}$ quarks in the QGP recombine to regenerate $J/\Psi$ (see Fig.~\ref{fig:f3}-(right)), miraculously making the observed $R_{AA}$ equal at SpS and RHIC c.m. energies. The good news is that such models predict $J/\Psi$ enhancement ($R_{AA}> 1$) at LHC energies, which would be spectacular, if observed.       
\subsection{Direct Photon Production}
\begin{figure}[!ht]
\begin{center}\vspace*{-2.0pc}
\begin{tabular}{cc}
\hspace*{-0.04\linewidth}\includegraphics[width=0.50\linewidth]{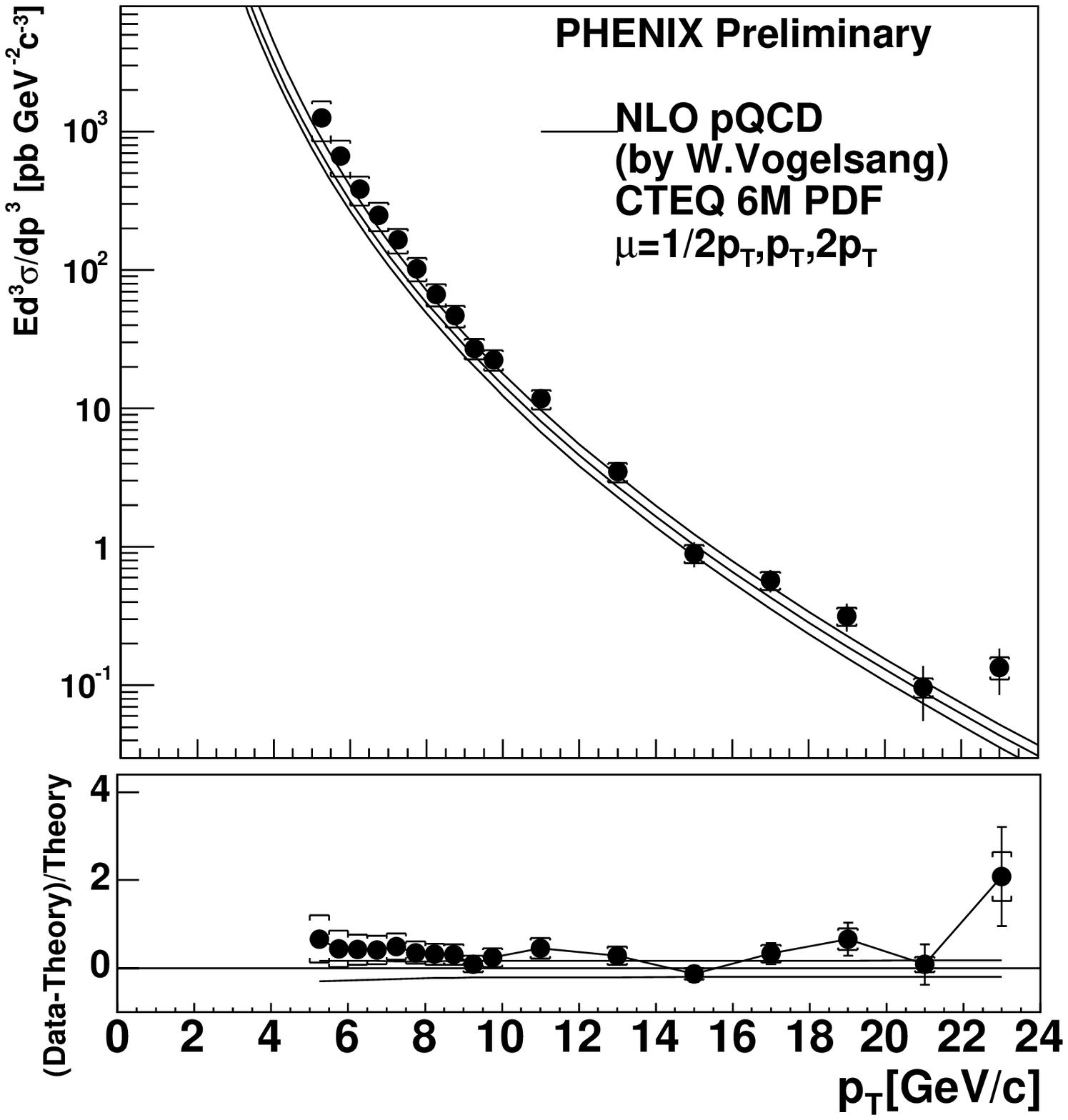} &
\hspace*{-0.06\linewidth}\includegraphics[width=0.50\linewidth]{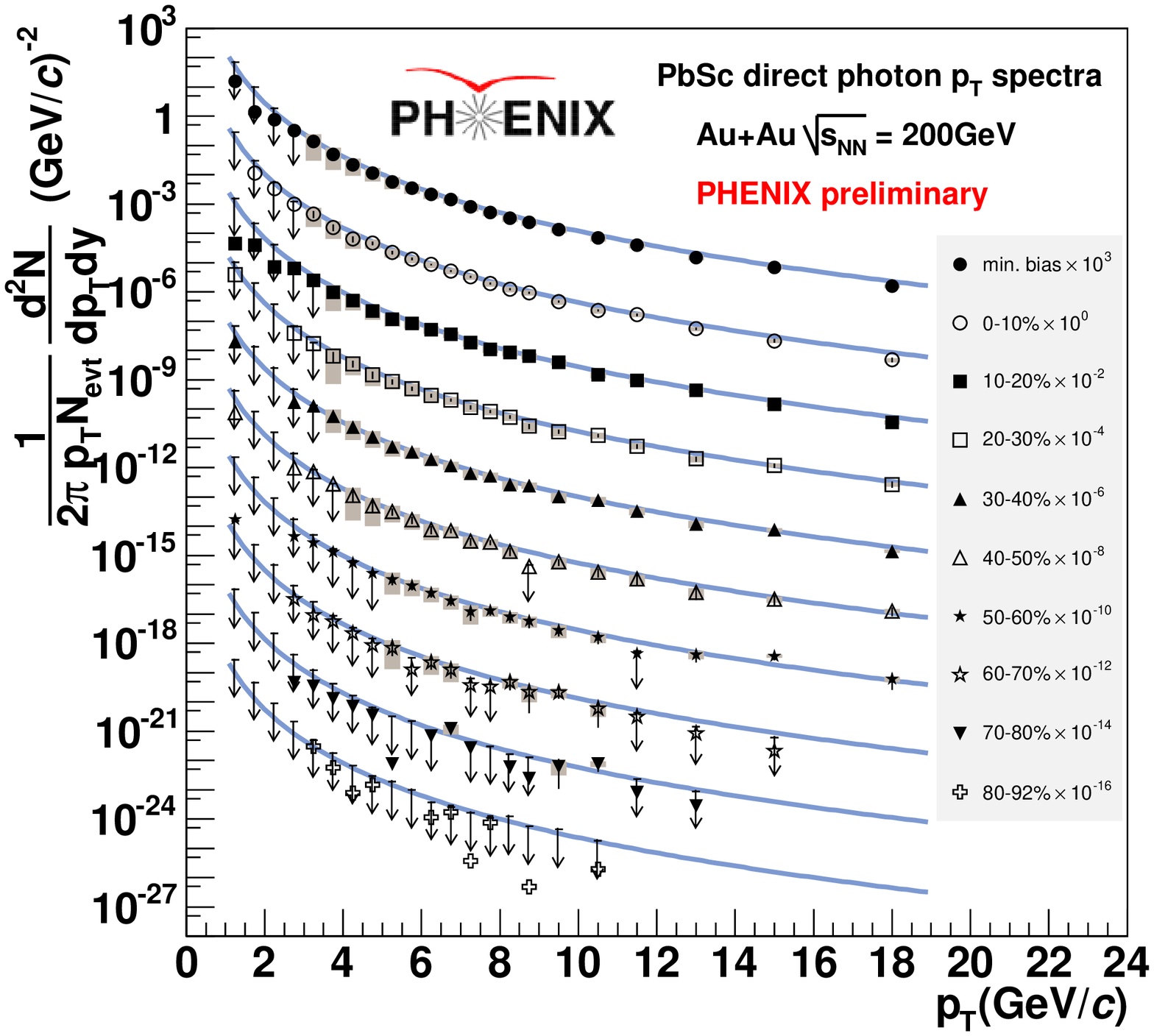} 
\end{tabular}\vspace*{-2.5pc}
\end{center}
\caption[]{(left) Invariant cross section for direct-$\gamma$ in p-p collisions at $\sqrt{s}=200$ GeV~\cite{QM06gamma}. (right) Invariant yield of direct-$\gamma$ in Au+Au as a function of centrality~\cite{DirG2}. Lines are QCD. }
\label{fig:f4}\vspace*{-1.5pc}
\end{figure}

\begin{figure}[!t]
\begin{center}
\begin{tabular}{c}
\includegraphics[width=0.80\linewidth]{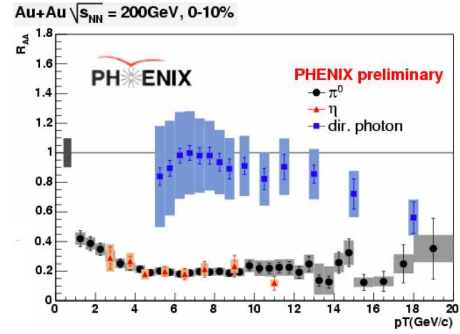} 
\end{tabular}
\end{center}\vspace*{-2.5pc}
\caption[]{$R_{AA}(p_T)$ for central (0-10\%) Au+Au collisions for direct-$\gamma$, $\pi^0$ and $\eta$~\cite{QM06gamma}.}
\label{fig:f5}\vspace*{-1.5pc}
\end{figure}

PHENIX has measured direct photon production in p-p and Au+Au collisions as a function of centrality at $\sqrt{s_{NN}}=200$ GeV over the range $4\leq p_T\leq 24$ GeV/c~\cite{DirG2} (Fig.~\ref{fig:f4}). 
The p-p measurements are in agreement with QCD and the Au+Au measurements agree with point-like scaling of the p-p cross section at all centralities, i.e. $R^{\gamma}_{AA}(p_T)\approx 1$ (see Fig.~\ref{fig:f5}). 
The direct-$\gamma$ are not suppressed for $p_T\leq 14$ GeV/c, while the $\pi^0$ and $\eta$ mesons, which are fragments of jets,  are suppressed by the same amount. This indicates that suppression is a final state effect on outgoing partons. For larger $p_T\sim 20$ GeV/c, the $\pi^0$ suppression is roughly constant at $R_{AA}\simeq 0.2$ while the direct-$\gamma$ appear become suppressed at a level approaching that of $\pi^0$ (with large systematic errors). If this were true, i.e. $R^{\pi^0}_{AA}=R^{\gamma}_{AA}$ for $p_T > 20$ GeV/c, it would indicate that the suppression in this higher $p_T$ range is not a final state effect (i.e. the energy loss becomes negligible compared to $p_T\sim 20$ GeV/c) and must be due to the structure functions. This is easy to understand for direct photons at mid-rapidity in minimum bias Au+Au for which:
\begin{equation}
R^{\gamma}_{AA}(p_T)\approx \left( \frac{F_{\rm 2A}(x_T)}{AF_{2p}(x_T)} \times \frac{g_{A}(x_T)}{Ag_{p}(x_T)}\right) \qquad ,
\end{equation}
where $F_{2p}(x)$ is the structure function from DIS and $g_p(x)$ is the gluon structure function. Unfortunately, there are no structure function measurement as a function of centrality in the 40 years of DIS~\cite{E665}. 
\subsection{Low $p_T$ direct photons}
\begin{figure}[!ht]
\begin{center}\vspace*{-2pc}
\begin{tabular}{cc}
\hspace*{-0.02\linewidth}\includegraphics[width=0.48\linewidth,height=0.5\linewidth]{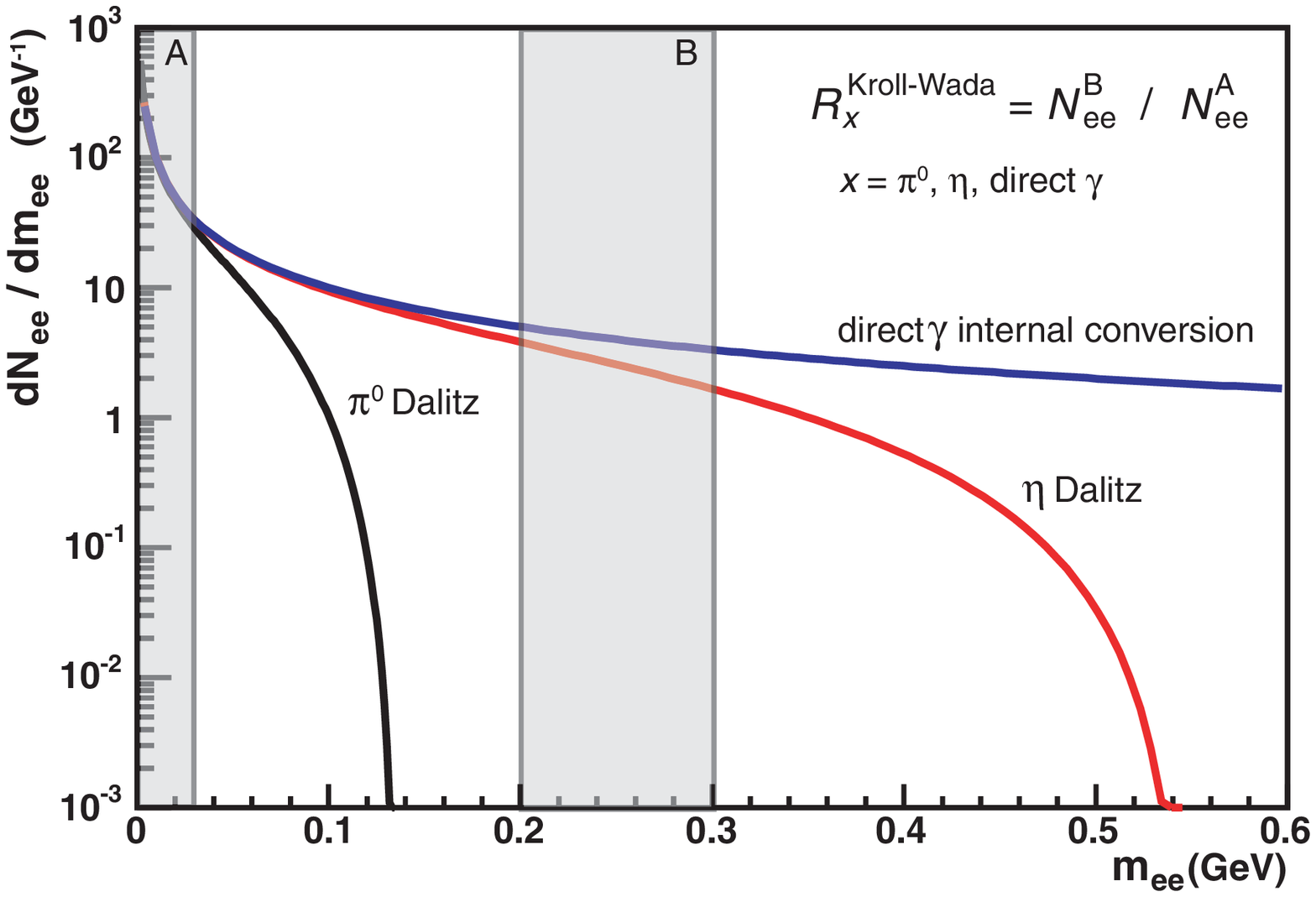} & \hspace*{-0.04\linewidth}
\includegraphics[width=0.48\linewidth]{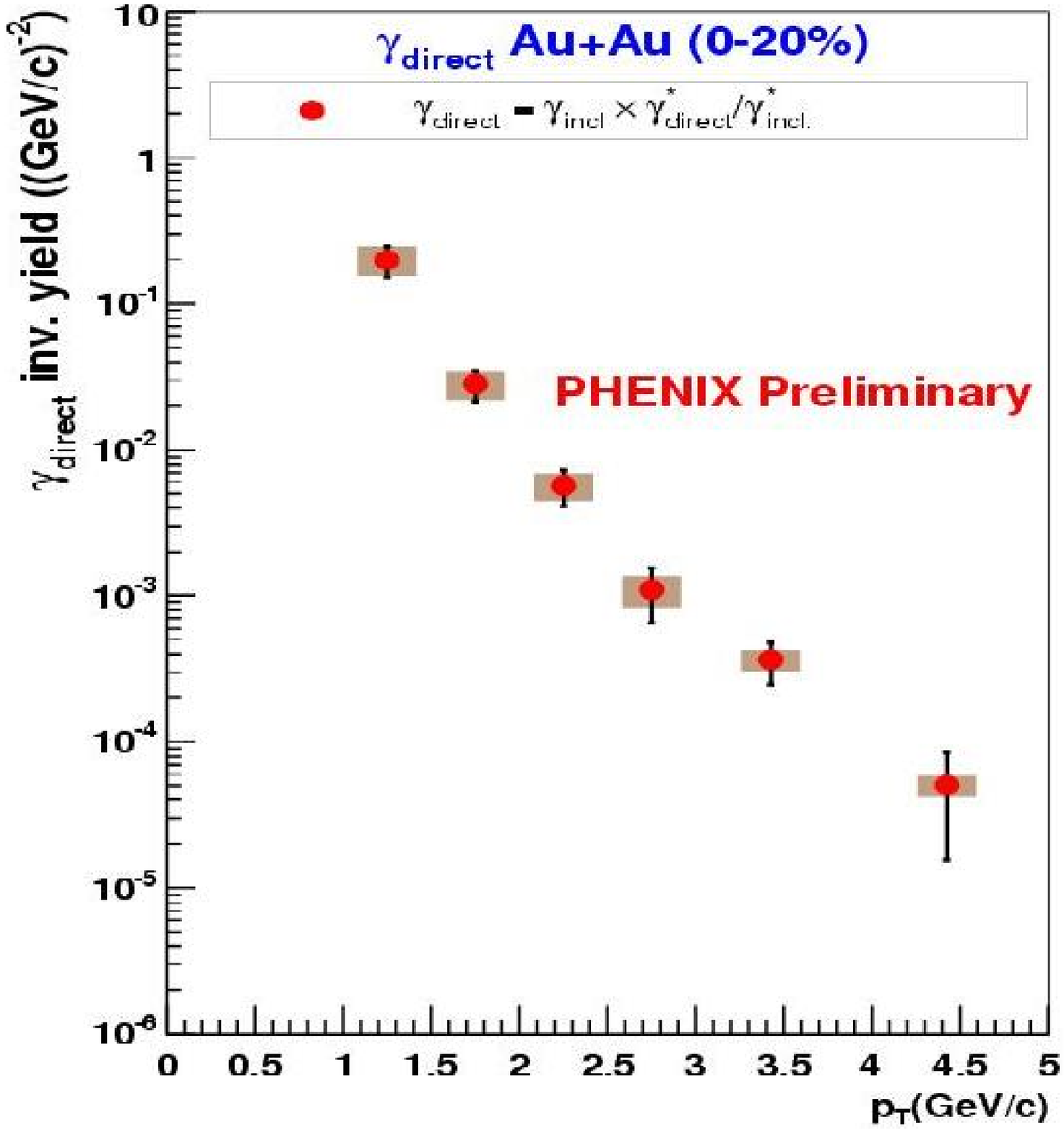} 
\end{tabular}
\end{center}\vspace*{-2.5pc}
\caption[]{(left) Kroll-Wada~\cite{KW} formula $dn_{ee}/dm_{ee}$ for $\pi^0$, $\eta\rightarrow \gamma\gamma$ and direct-$\gamma$ internal conversions; PHENIX Invariant yield of direct-$\gamma$ in central (0-20\%) Au+Au collisions via internal conversions~\cite{YAQM05}. }
\label{fig:f6}\vspace*{-2pc}
\end{figure}

Studies of direct photons for $p_T\leq 3$ GeV/c are difficult for two reasons: the $\pi^0$ background becomes prohibitive; the theory tends to diverge. However, the rate is high, so that measurement by internal conversion becomes practical. As shown in Fig.~\ref{fig:f6}-(left), internal conversions from $\pi^0$ and $\eta\rightarrow \gamma\gamma$ are kinematically suppressed, and vanish above the rest mass, while direct-$\gamma$ from $g+q\rightarrow \gamma+q$ are not kinematically suppressed for $m_{ee} \ll p_T$. Thus, the direct-$\gamma$ component can be extracted with much less (even no) background by making measurements as a function of $m_{ee}$. The direct-$\gamma$ spectrum for $1\leq p_t\leq 5$ GeV/c in central (0-20\%) Au+Au collisions was derived by this method (Fig.~\ref{fig:f6}-(right)) and appears to be exponential for $p_T\leq 3$ GeV/c. This would be the smoking gun for thermal photons, if the direct-$\gamma$ spectrum at low $p_T$ in p-p collisions were gaussian as for Drell-Yan rather than exponential as for $\pi^0$. Unfortunately, this has never been measured, and we are working on it.  

\subsection{Charm via direct $e^{\pm}$.}
\begin{figure}[!ht]
\begin{center} 
\begin{tabular}{cc}
\hspace*{-0.04\linewidth}\includegraphics*[width=0.53\linewidth]{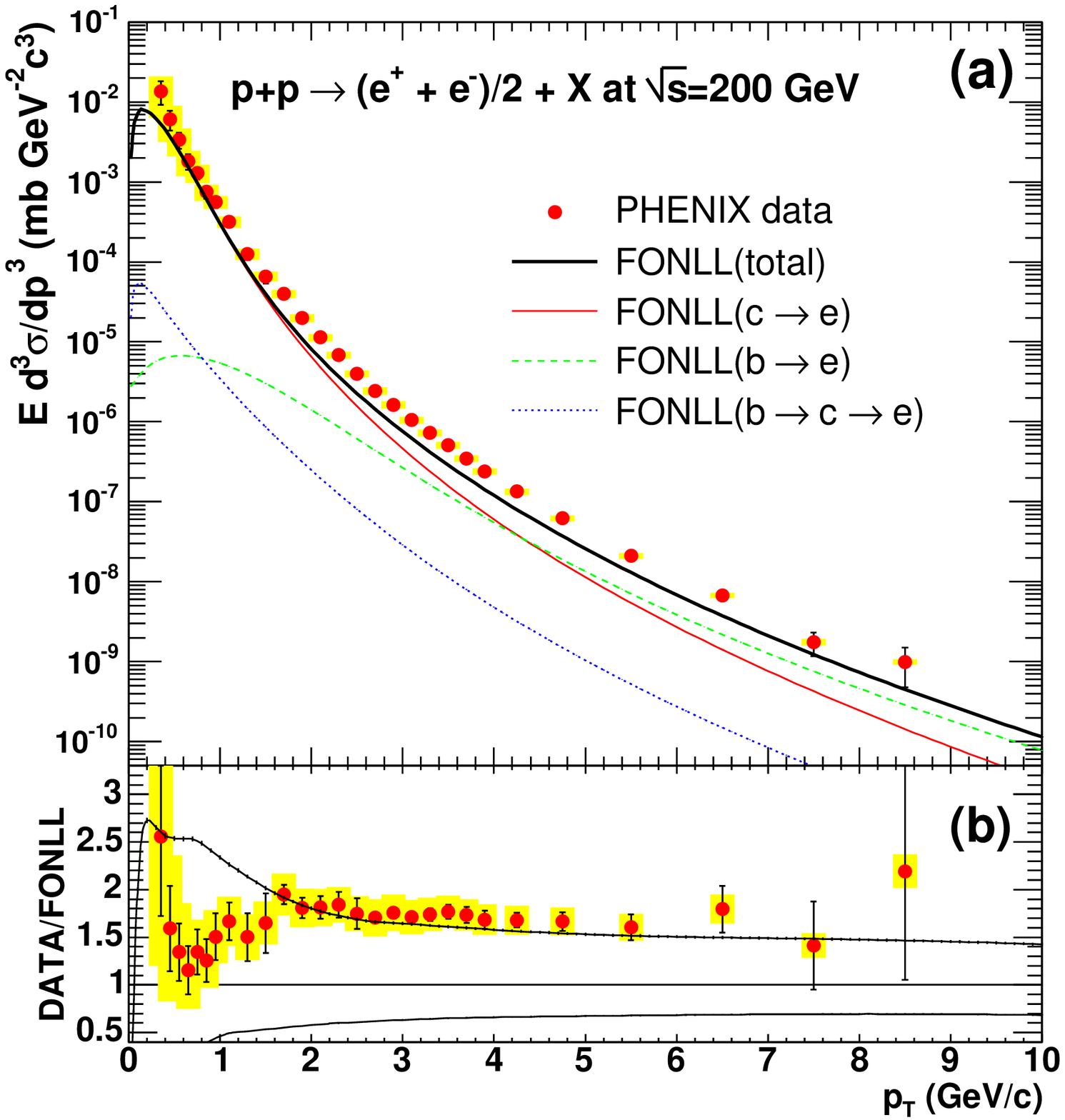} & 
\hspace*{-0.08\linewidth}\includegraphics[width=0.51\linewidth]{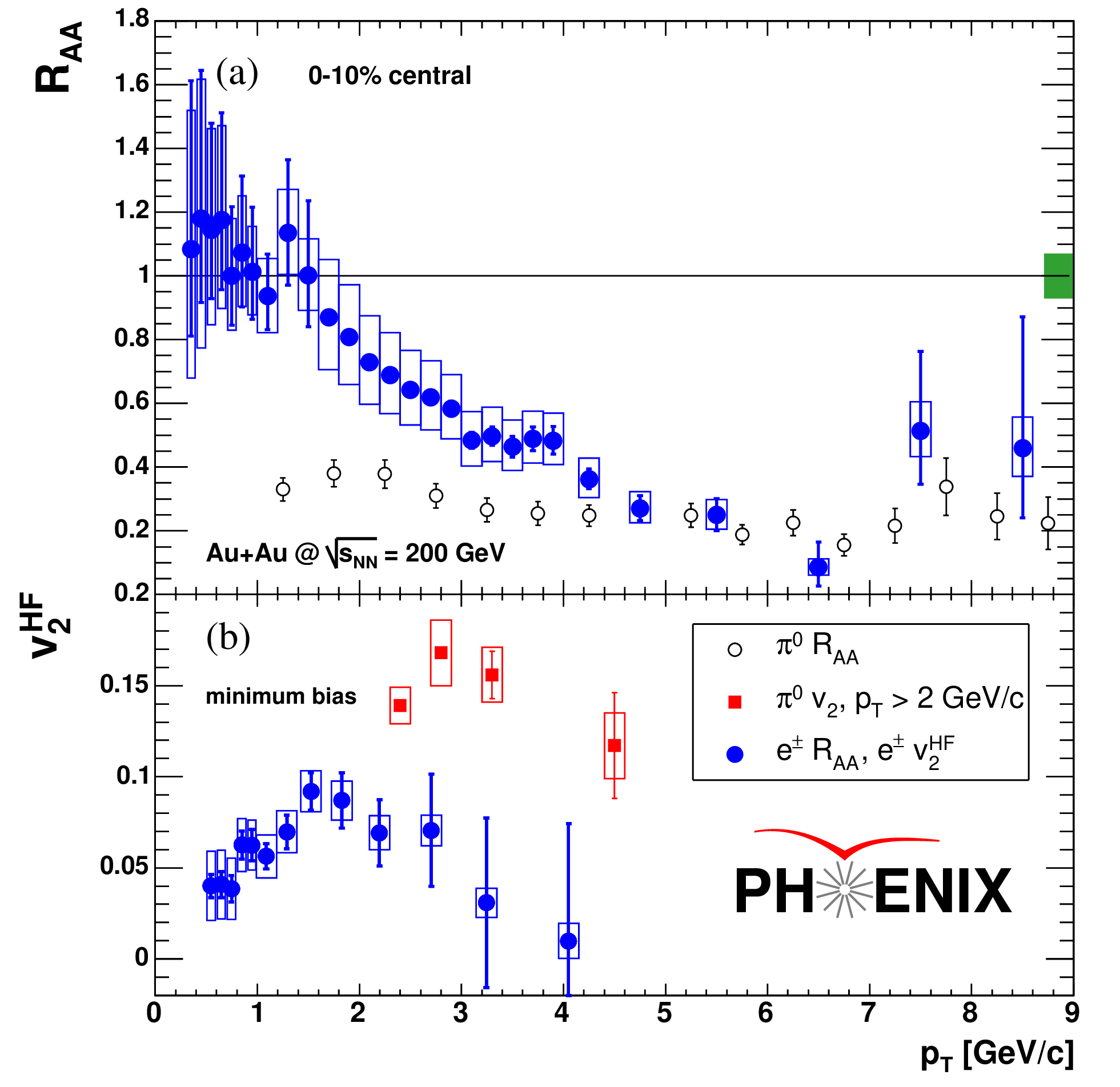} 
\end{tabular}
\end{center}\vspace*{-2.5pc}
\caption[]{(left) Invariant cross section of direct $e^{\pm}$ in p-p collisions ~\cite{PXPRL97e}. (right) a) $R_{AA}(p_T)$ of direct $e^{\pm}$ in central (0-10\%) Au+Au collisions ccompared to $R^{\pi^0}_{AA}(p_T)$~\cite{PXPRL98e}; b) $v_2$ for $e^{\pm}$ and $\pi^0$ see Ref.~\cite{PXPRL98e} for details.}
\label{fig:f7}\vspace*{-1.5pc}
\end{figure}

PHENIX, which was specifically designed for this purpose, has measured direct $e^{\pm}$ in both p-p and Au+Au collisions out to $p_T\sim 9$ GeV/c. The p-p result is in agreement with a QCD calculation of $c$ and $b$ quarks as the  source of the direct $e^{\pm}$ (Fig.~\ref{fig:f7}-(left)), but a totally surprising result is that the direct $e^{\pm}$ (and thus the heavy quarks) are suppressed the same as $\pi^0$ in the range $4\leq p_T\leq 9$ GeV/c where $b$ and $c$ contributions are roughly equal (Fig.~\ref{fig:f7}-(right)a). This is presently not understood but seems to argue against a naive picture of radiative energy loss to explain the suppression. 
\subsection{Jets via 2-particle correlations.}
\begin{figure}[!ht]
\begin{center}\vspace*{-1.5pc}
\begin{tabular}{cc}
\begin{tabular}[b]{c}
\hspace*{-0.03\linewidth}\includegraphics[width=0.48\linewidth]{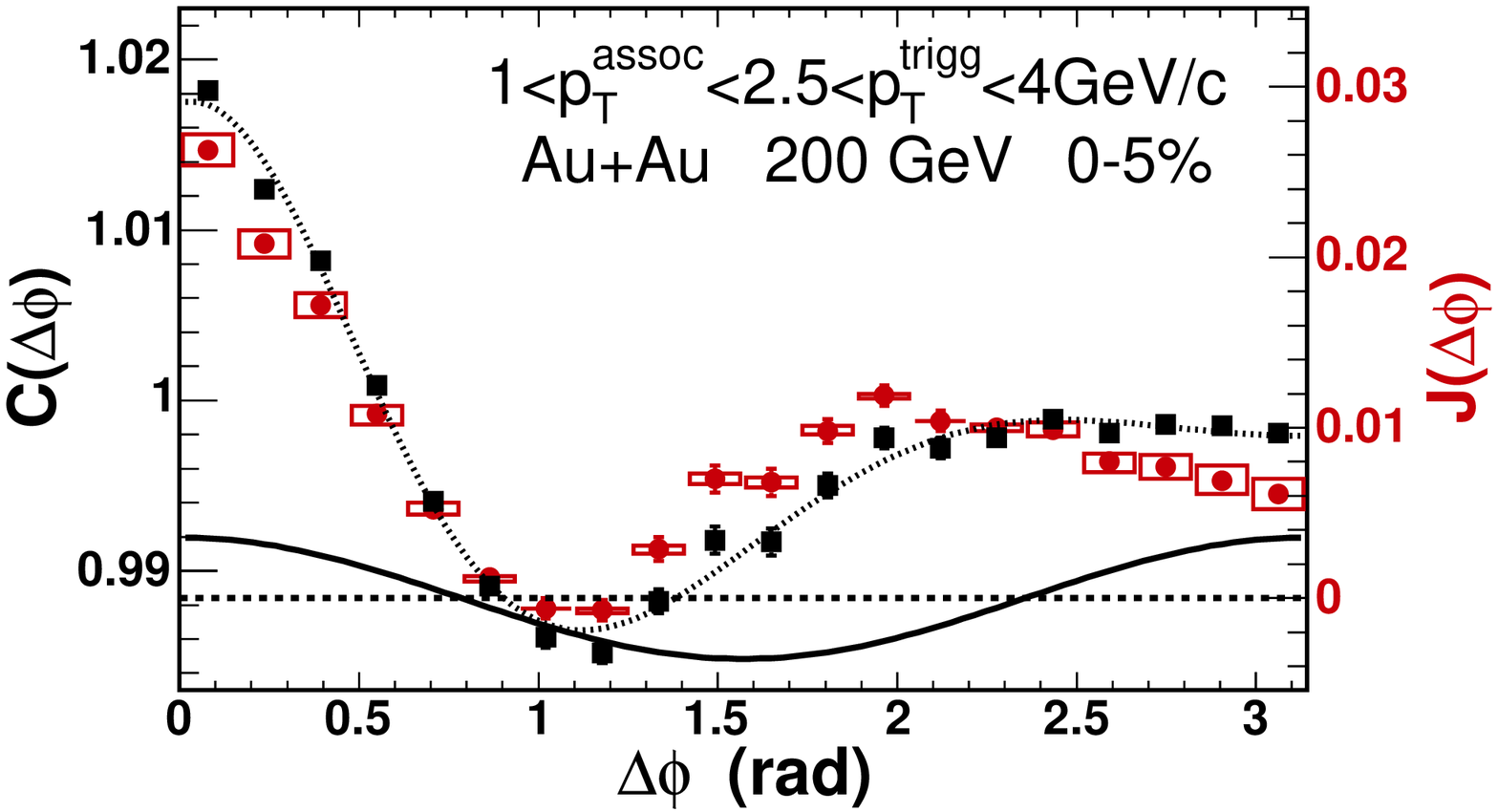}\cr
\hspace*{-0.03\linewidth}\includegraphics[width=0.41\linewidth]{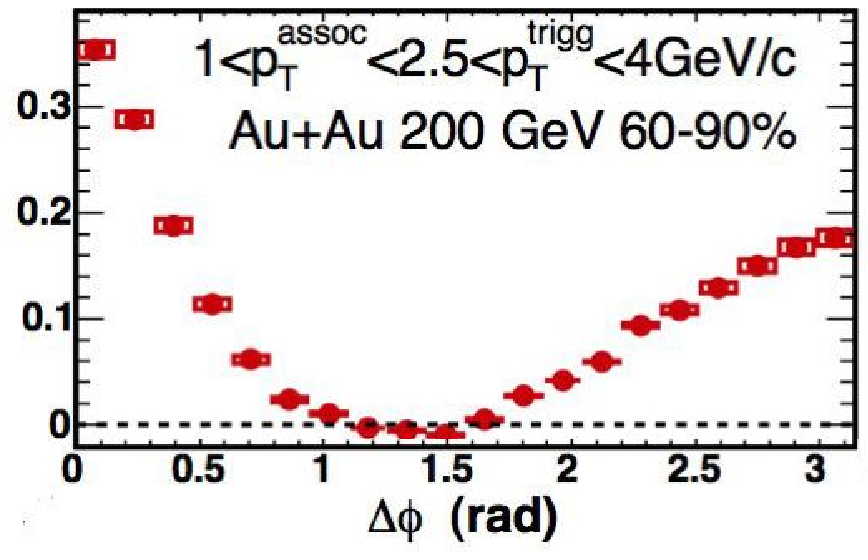}\hspace*{0.03\linewidth}
\end{tabular} 
\hspace*{-0.03\linewidth}\includegraphics[width=0.48\linewidth]{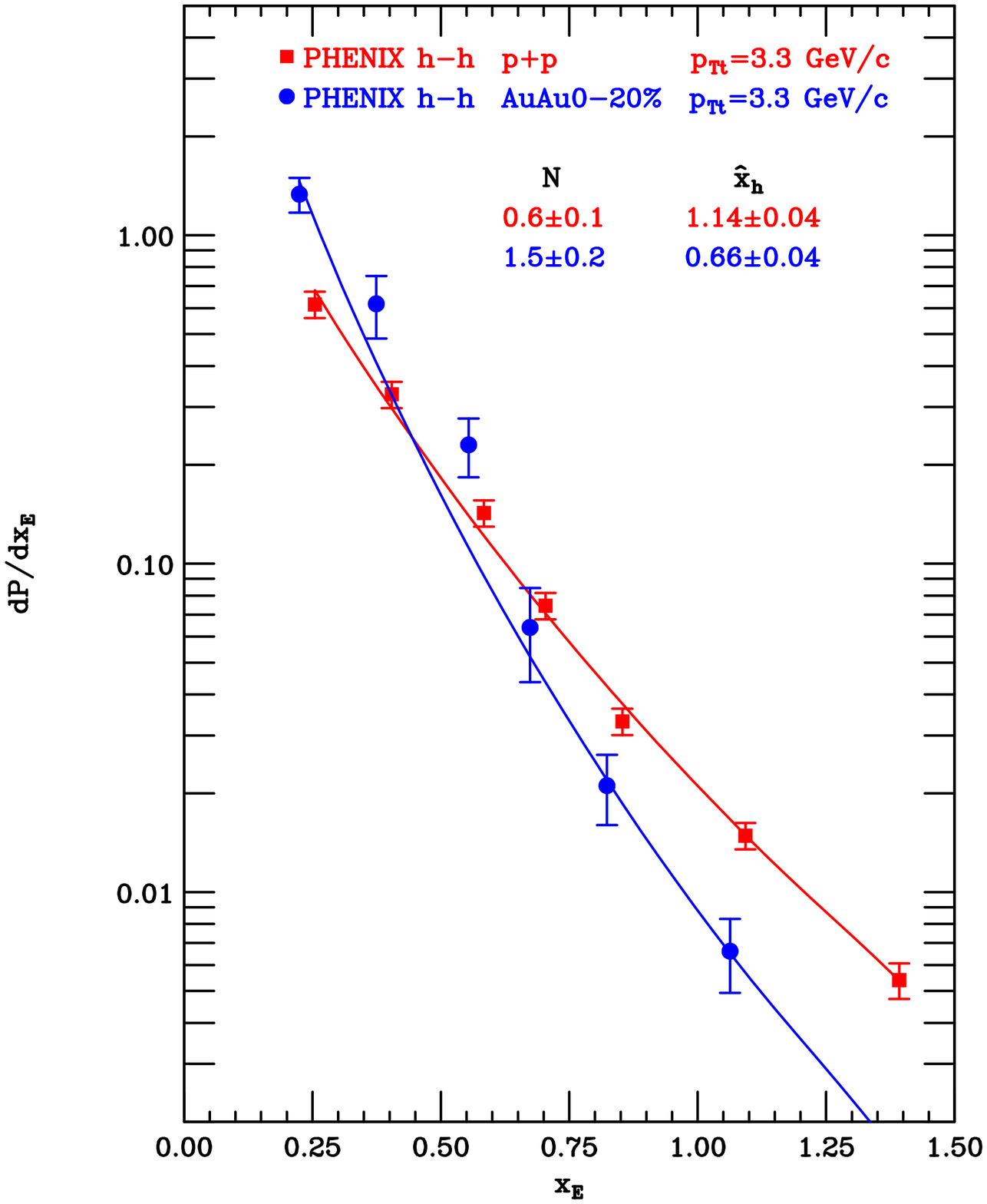} 
\end{tabular}
\end{center}\vspace*{-2.7pc}
\caption[]{(left) Azimuthal correlation $C(\Delta\phi)$ of $h^{\pm}$ with $1\leq p_{T_a}\leq 2.5$ GeV/c with respect to a trigger $h^{\pm}$ with $2.5\leq p_{T_t} \leq 4$ GeV/c in Au+Au: (top) central collisions, where the line with data points indicates $C(\Delta\phi)$ before correction for the azimuthally modulated ($v_2$) background, and the other line is the $v_2$ correction which is subtracted to give the jet correlation function $J(\Delta\phi)$ (data points); (bottom)-same for peripheral collisions. (left) $x_E\approx p_{T_a}/p_{T_t}$ distribution for the Au+Au-central data compared to p-p.}
\label{fig:f8}\vspace*{-1.5pc}
\end{figure}

The huge multiplicity in central A+A collisions, roughly $A$ times the multiplicity in p-p collisions, makes jet reconstruction difficult if not impossible at RHIC. However, two-particle correlations work well, with the major complication being the event-anisotropy ($v_2$) with respect to the reaction plane defined by the impact parameter vector in Au+Au collisions (Fig.\ref{fig:f8}-(left)).  
Even before the $v_2$ subtraction, the away-side distribution in central collisions is much wider than in peripheral collisions (which is similar to p-p). After the $v_2$ correction, a double peak structure $\sim \pm 1$ radian from $\pi$ is evident, with a dip at $\pi$ radians. This may indicate the reaction of the medium to the away-parton and is under active study. We discovered in this analysis~\cite{ppg029} that the yield in the away-side peak plotted as a function of the scaled variable, $x_E\approx p_{T_a}/p_{T_t}$, does not measure the fragmentation function, as believed since ISR studies~\cite{FFF}. For an exponential fragmentation function ($e^{-bz}$) and a power law invariant parton distribution ($p_T^{-n}$) the $x_E$ distribution is~\cite{ppg029}:
\begin{equation}
\left . {dP\over dx_E}\right |_{p_{T_t}}=N (n-1) {1\over \hat{x}_h} {1 \over{(1+x_E/\hat{x}_h})^n}
\label{eq:xEdist} 
 \end{equation}
 where $\hat{x}_h$ is the ratio of the transverse momenta of the away-side  parton to the trigger parton and $N$ is a normalization factor. Fig.~\ref{fig:f8}-(right) shows fits of Eq.~\ref{eq:xEdist} to the $x_E$ distributions in p-p and Au+Au central collisions. The steeper Au+Au distribution indicates a smaller $\hat{x}_h$ in Au+Au collisions than in p-p collisions, evidence for energy loss of the away-side parton in the medium produced in Au+Au central collisions.   

\end{document}